\documentclass[a4paper]{jpconf}
\usepackage{graphicx}
\usepackage{type1cm}
\usepackage{amssymb}
\newcommand{\degree}{\ensuremath{^\circ}}

\DeclareFontShape{OT1}{cmtt}{bx}{n}
{
<5><6><7><8><9><10><10.95><12><14.4><17.28><20.74><24.88>cmttb10
}{}

\begin{document}
\title{The Fluorescence Detector of the Pierre Auger Observatory}

\author{Petr Ne\v{c}esal for the Pierre Auger Collaboration}

\address{Fyzik\'{a}ln\'{i} \'{u}stav AV \v{C}R, v. v. i., Na Slovance 2, Praha 8, 182 21, Czech Republic and
Observatorio Pierre Auger, Av. San Martin Norte 304, 5613 Malarg\"{u}e, Argentina;
http://www.auger.org/archive/authors\underline{ }2010\underline{ }08.html} %\url{http://www.auger.org/archive/authors_2010_08.html}} %\href{http://www.auger.org/archive/authors_2010_08.html}{http://www.auger.org/archive/authors\underline{ }2010\underline{ }08.html}"
%; http://www.auger.org/auger-authors.pdf

\ead{necesal@fzu.cz}

\begin{abstract}
The Pierre Auger Observatory is a facility designed for the study of ultra-high energy cosmic rays. The Observatory combines two different types of detectors: a surface array of 1600 water Cherenkov stations placed on a 1.5 km triangular grid covering over 3000 km$^2$; and a fluorescence detector of 24 telescopes located in 4 buildings at the perimeter of the surface array. The fluorescence telescopes, each consisting of 440 photomultipliers, collect the ultraviolet light produced when the charged secondary particles in an air shower excite nitrogen molecules in the atmosphere. Because the intensity of the nitrogen fluorescence is proportional to the energy deposited in the atmosphere during the air shower, the air fluorescence measurements can be used to make a calorimetric measurement of the cosmic ray primary energy. Showers observed independently by the surface array and fluorescence telescopes, called hybrid events, are critical to the function of the Observatory, as they allow for a model-independent calibration of the surface detector. In this paper I describe the detector and the most important measurements.
\end{abstract}

\section{Introduction}
Ultra-high energy cosmic rays (UHECRs) are particles coming to the atmosphere of the Earth with energies that can exceed $10^{20}$ eV. %These energies are much higher than achievable at the largest Earth's accelerators. 
While the first observation of cosmic rays (CRs) was made one hundred years ago, origins of these particles is not yet understood. %Cosmic ray research remains very interesting even one hundred years after their discovery. 
%Particles with energy much higher than achievable at the largest Earth's accelerators carry messages from their sources. 
Much of the reason is because of the very low flux of 
UHECRs at Earth. The differential flux of UHECRs is steeply falling with energy, and can be described by a power law
\begin{equation}
\label{eqFlux}
\frac{dN}{dE} \propto E^{-\gamma}
\end{equation}
with spectral index $\gamma$ between 2.6 and 3.2.  
One can find two significant points in the energy spectrum -- at $4\times 10^{15}$ eV there is so called 'knee' where $\gamma$ changes from 2.7 to 3.1 \cite{ralfE} and known 'ankle' can be found at $4\times 10^{18}$ eV with another spectral index change from 3.3 to 2.6~\cite{spectrum}. Processes and mechanisms of achieving ultra-high energies are still undiscovered. So the main questions which people want to answer concern composition, energy spectrum and sources of UHECRs.

Primary CRs hitting the Earth's atmosphere interact with air molecules and fragment.The secondary particles produced in this collision will undergo further hadronic and electromagnetic interactions, ultimately producing an extensive air shower cascade. %Secondary particles originating in this interaction impact another air atomic nuclei and cause interactions again. A number of all secondary particles is growing. A successive cascade of hadronic and electromagnetic interactions initiates Extensive Air Shower (EAS).
As the air shower develops in the atmosphere the number of secondary particles increases and their average energy falls. Due to energy losses and decays, the shower will reach a maximum size, and then number of particles in the shower will begin to decrease. However, there are still enough energetic particles in a vertical air shower which can hit ground and can be detected by ground detectors. With increasing zenith angle of the primary particle an air shower must propagate through larger atmospheric mass and therefore only less particles survive at the ground level. 

There are two possible types of UHECR detection -- direct and indirect methods. Direct methods can be performed by means of satellites and balloons, but cannot considerably measure UHECR at energies larger than $10^{14-15}$ eV because of small statistics (Eq. \ref{eqFlux}). On the other hand indirect methods require sufficient energy $(E \gtrsim 10^{15})$ eV to develop shower and large detector area is necessary. Indirect methods are based on the measurement of secondary particles that reach the ground in array of Cherenkov detectors, scintillators or muon detectors or on the collection of the isotropic fluorescence light. Secondary charged particles in the shower excite nitrogen molecules, which then isotropically emit fluorescence light into several bands between 300 and 420 nm. These photons can be detected by fluorescence telescopes.
%and enable calorimetric measurement.

\begin{figure}[ht]
\begin{minipage}{21.5pc}
\includegraphics[width=22pc]{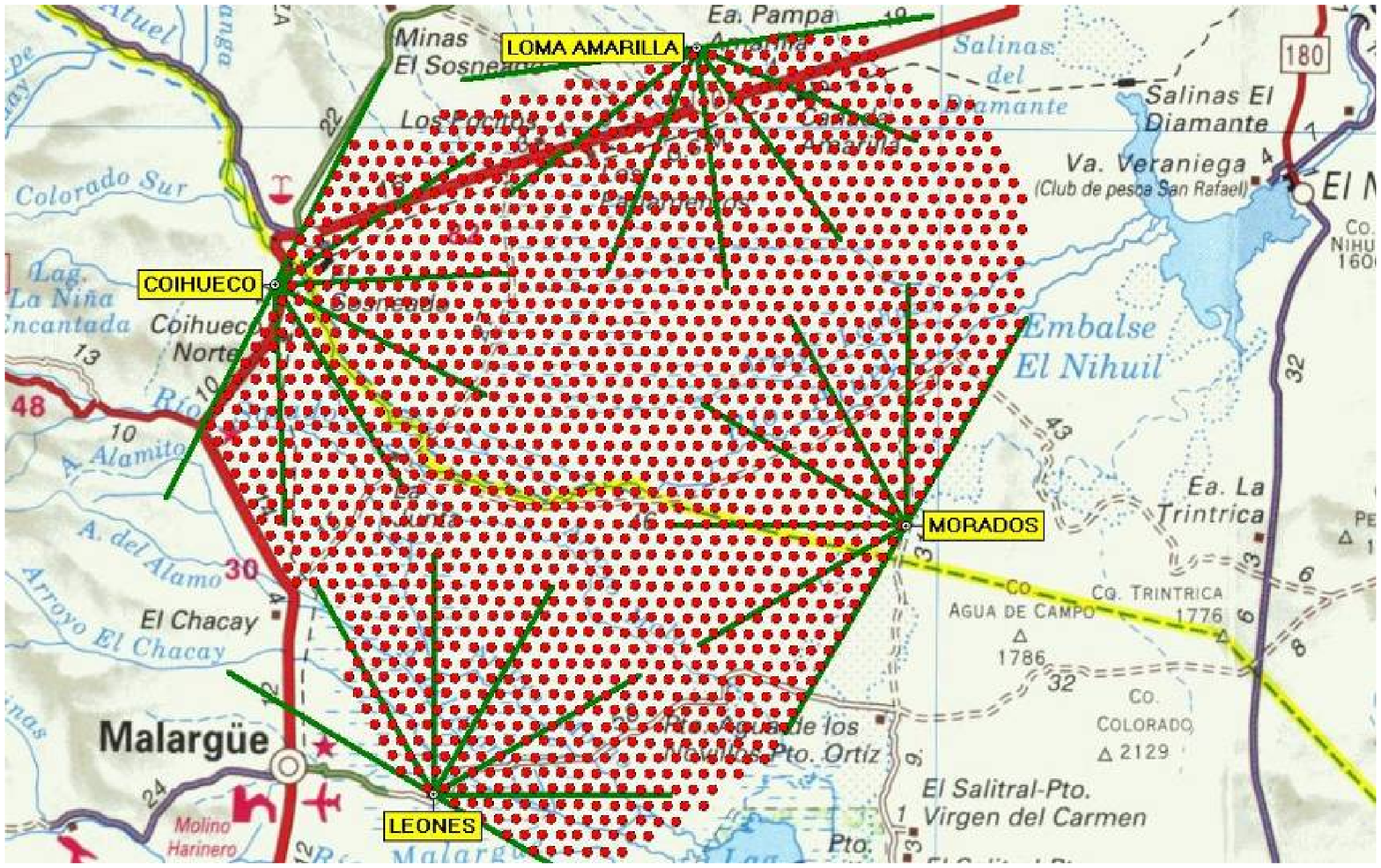}
\caption{\label{figArray}The detector layout of the Pierre Auger Observatory. Red dots indicate SD stations which are overviewed by 4 FD buildings. There are 6 fluorescence telescopes in each building. Green lines indicate field of view of each telescope.}
\end{minipage}\hspace{2pc}%
\begin{minipage}{14pc}
\includegraphics[width=14pc]{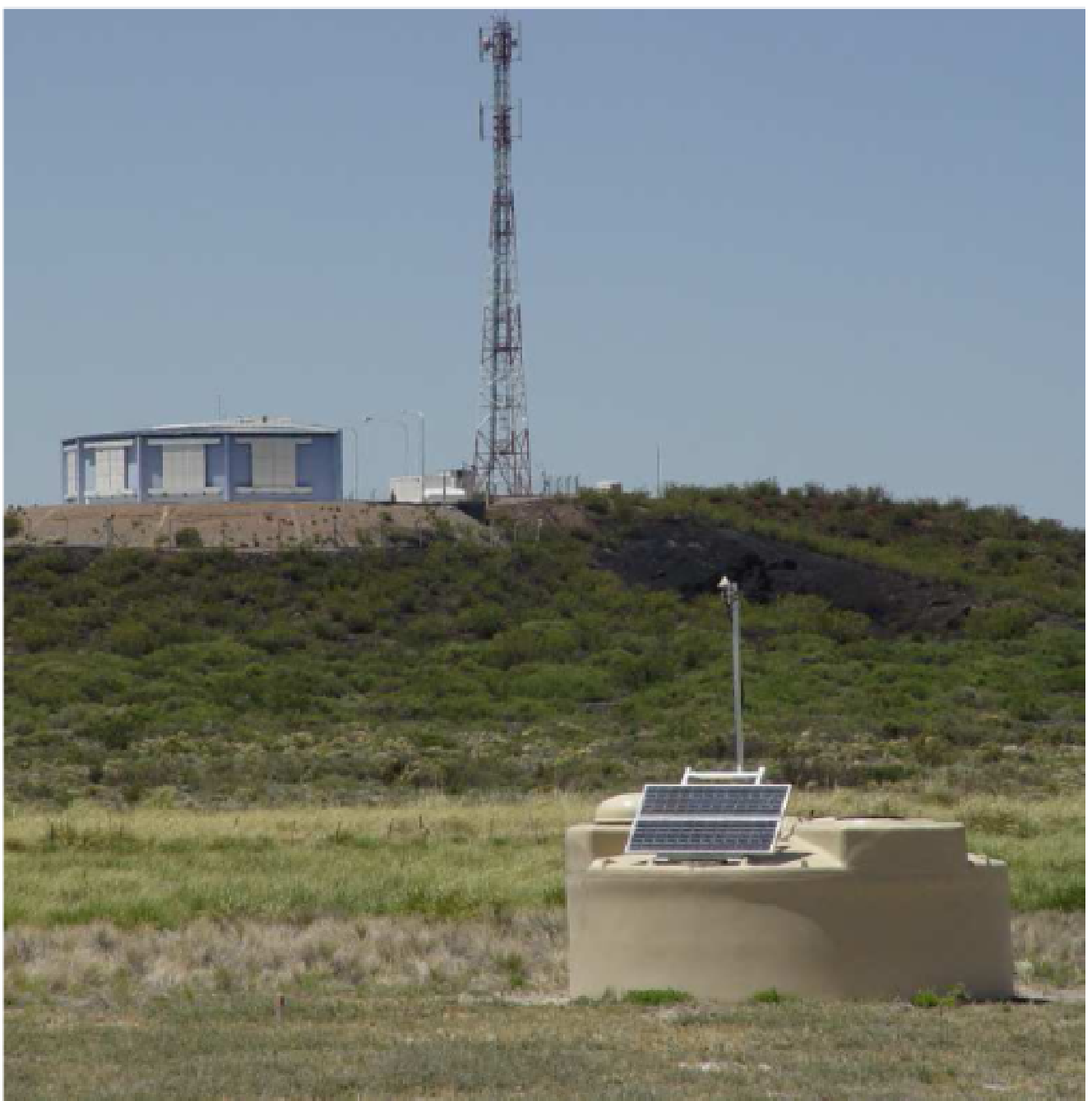}
\caption{\label{figSdFd}A picture of an SD station and FD building. Each station has its own solar panel with battery and communication antenna. There is also the antenna nearby the FD building.}
\end{minipage}
\end{figure} 

\section{Pierre Auger Observatory}
The Pierre Auger Observatory \cite{augerOb} is a facility studying UHECRs by means of two type of detectors: a surface detector (SD) which is composed of 1600 water Cherenkov stations and a fluorescence detector (FD) that comprises 24 fluorescence telescopes. The southern site of the Observatory is located in province Mendoza in Argentina near the city Malarg\"{u}e ($69\degree$~W, $35\degree$~S, 1420 m a.s.l.), the northern site is intended to be built in Colorado, USA. The layout of the southern site detectors is shown in Fig. \ref{figArray}, and the picture of an SD station and an FD building is shown in Fig. \ref{figSdFd}. Stations are deployed on a triangular grid with 1.5 km spacing. They cover an area of over 3000 km$^2$ and they have nearly 100 $\%$ duty cycle. There are 3 photomultipliers in pure water in each station. 

%Tanks send data to the nearest communication tower. They are powered by solar-charged batteries.

The Pierre Auger Observatory is designed to measure UHECRs with energy above $\sim 10^{18}$~eV. However, there is also a low energy extension of SD station infill with plastic scintillators beneath the stations to detect muons (AMIGA \cite{amiga_heat}) and a set of 3 fluorescence telescopes with high elevation called HEAT \cite{amiga_heat}. An extensive program of atmospheric monitoring is essential part of the Observatory not only to control uncertainties concerning FD.

\section{Fluorescence Detector}
\begin{figure}[ht]
\begin{minipage}{18.5pc}
\includegraphics[width=18pc]{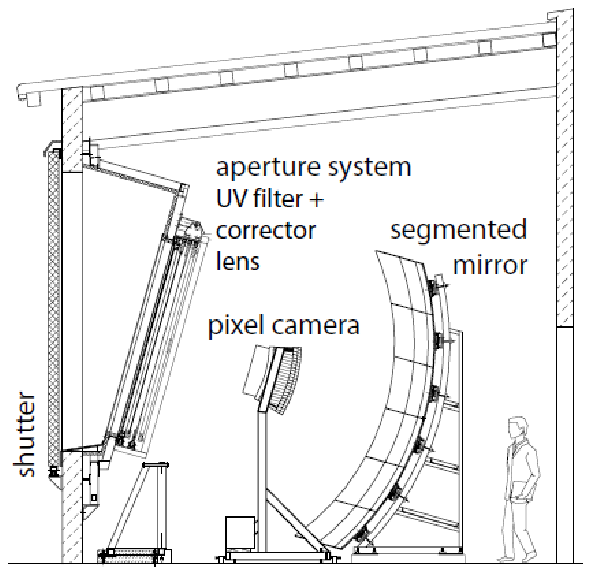}
\caption{\label{figFdLayout}A scheme of telescope geometry. Light comes through the aperture, covered by a UV filter and a corrector ring, and it is reflected by a segmented mirror to the pixel camera with 440 photomultipliers.}
\end{minipage}\hspace{2.0pc}%
\begin{minipage}{17.0pc}
\includegraphics[width=15pc]{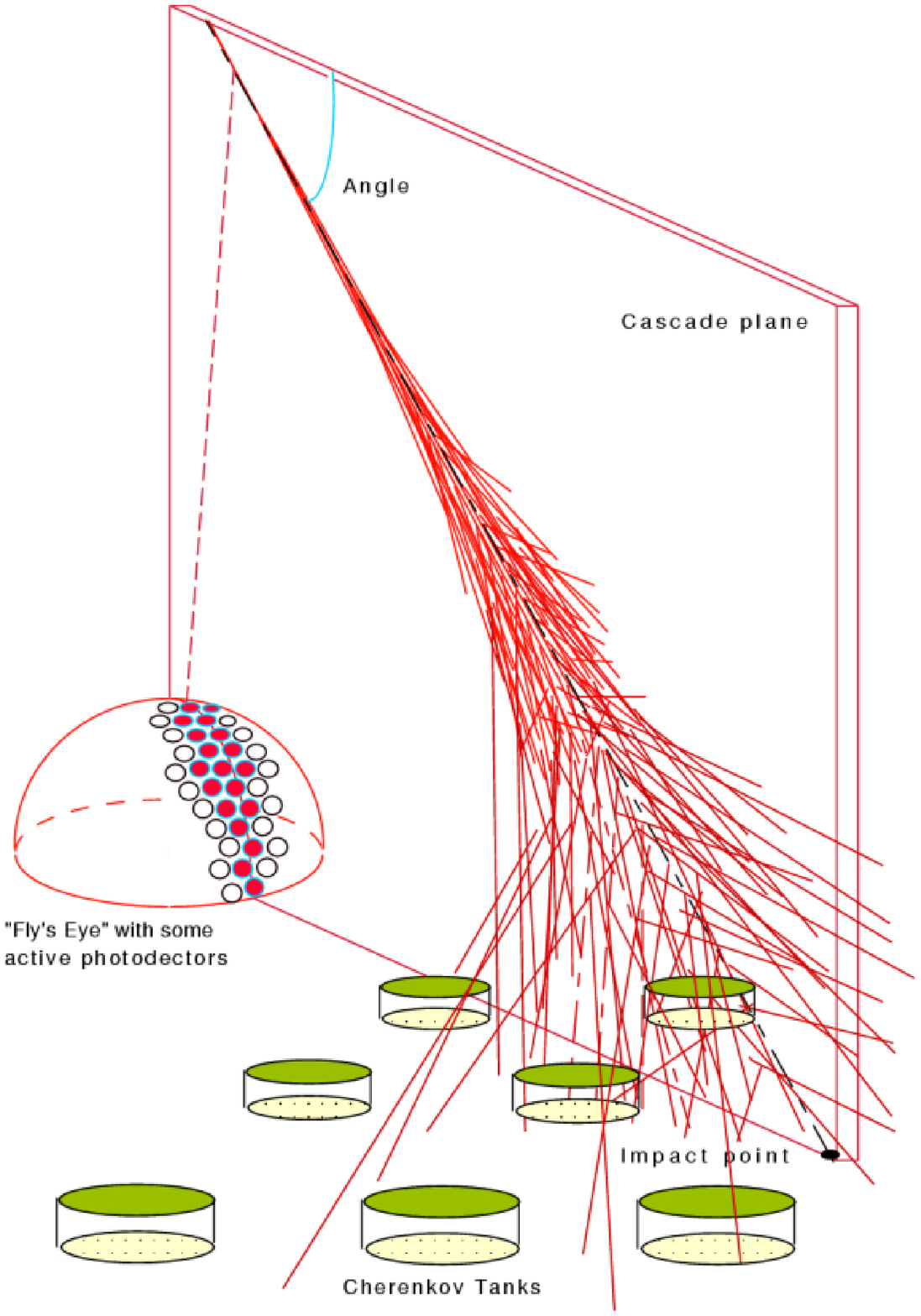}
\caption{\label{figHybridDetection}A cartoon of the hybrid detection technique. Charged particles at ground level are detected in SD stations while fluorescence light is collected in FD.}
\end{minipage}
\end{figure}

There are 4 buildings with 6 fluorescence telescopes each that overlook the area of SD stations (Fig. \ref{figArray}). Each telescope uses Schmidt optics and consists of a wide-angle, segmented spherical mirror, a spherical focal plane, a UV 300-410 nm passband filter, and a refractive corrector ring at the aperture of the telescope. The telescope field of view is $30\degree$ (in azimuth) x $28\degree$ (in elevation) so each building has $180\degree$ azimuth range (the 3 high-elevation telescopes observe $30\degree$ to $60\degree$ in elevation). There are 440 photomultipliers in the focal plane which collect reflected light. The longitudinal profile of a shower is thus measured as an image of active PMT pixels along the shower axis. The layout of the FD geometry is shown in Fig. \ref{figFdLayout}. 

There are several trigger levels. The first level digitizes signals from an analog board in each pixel at 10 MHz. The second level trigger is also an internal trigger to search for track segments of at least five active pixels. Every 100 ns, a scan over the full camera is performed and the triggered pixels are searched for track-like patterns. The third level trigger is a software algorithm that rejects events caused by lightning, muons which impact the focal plane, or randomly trigged pixels.

%... energy by integrating longitudinal profile, FD...
FD telescopes are calibrated to find the proper conversion between digitized counts and the true light flux (in photons). This calibration is performed in several steps using absolute and relative methods. The absolute calibration uses a calibrated light source (known as a 'drum') mounted at the telescope aperture \cite{fdPaper}. Using the drum, the light flux at each pixel is known and the response is measured. In addition to the drum calibration, vertical laser shots at wavelengths 337 and 355 nm \cite{roberts, knapik} are used as an independent calibration method.
Between absolute calibrations (which occur several times per year), the relative response of each fluorescence camera is recorded using light pulses from LEDs and xenon flashers. The relative calibration occurs before and after each night of observations.

%\\doplnit nejakou vetu, ze se rekonstruuje nejprve shower axis z casu z %fotonasobicu a taky, ze je treba odecist cherenkov - viz fd paper.

\subsection{Hybrid Detection Technique}
\begin{figure}[htb]
\begin{minipage}{19.0pc} %20
\includegraphics[width=20pc]{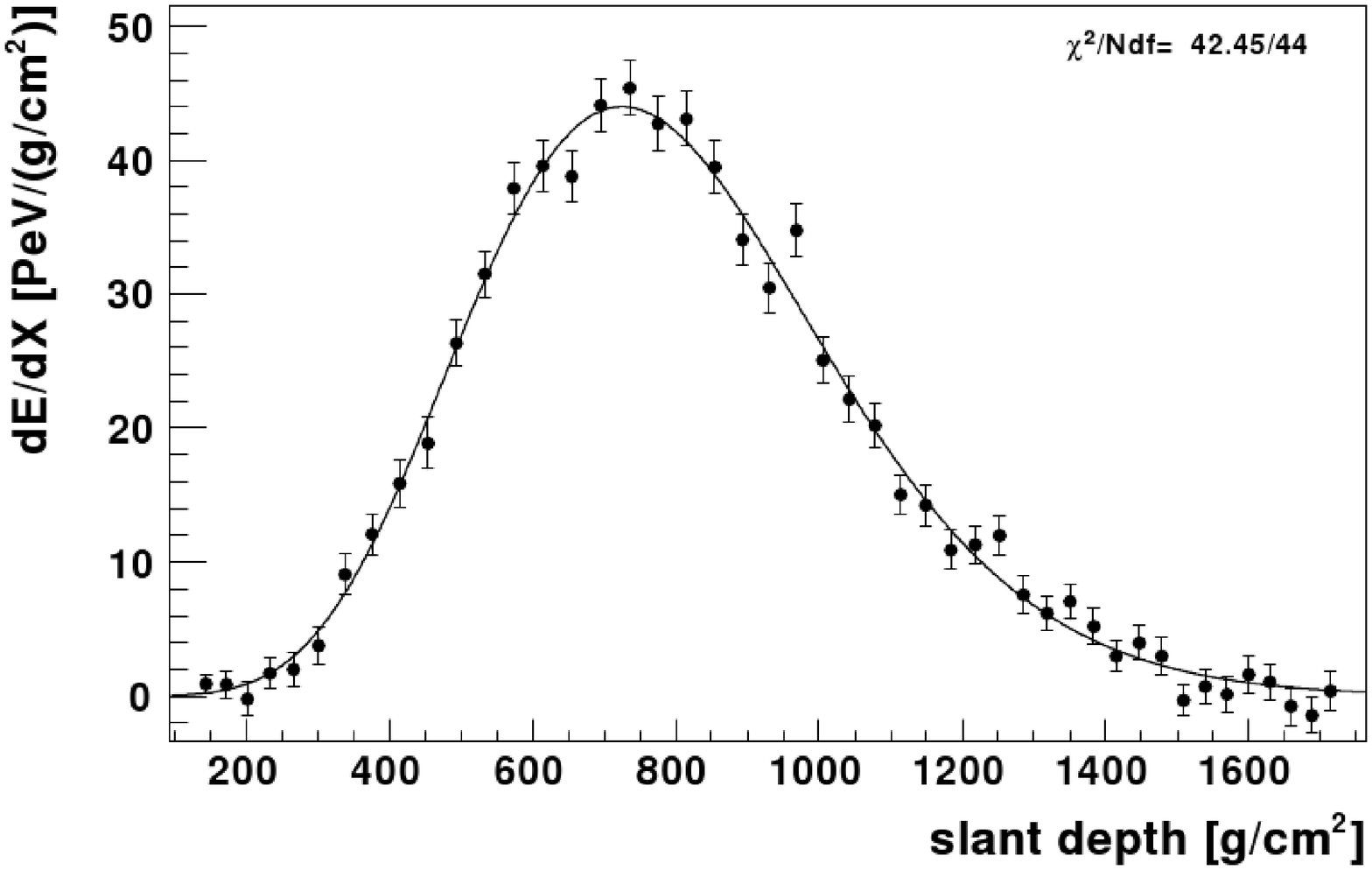}
\caption{\label{figLogProf} Longitudinal profile of energy deposit reconstructed from measured light by FD telescope. The line represents a Gaisser-Hillas function fitted to the measured profile.}
\end{minipage}\hspace{1pc}%
\begin{minipage}{17.5pc} %16.5
\includegraphics[width=17.5pc]{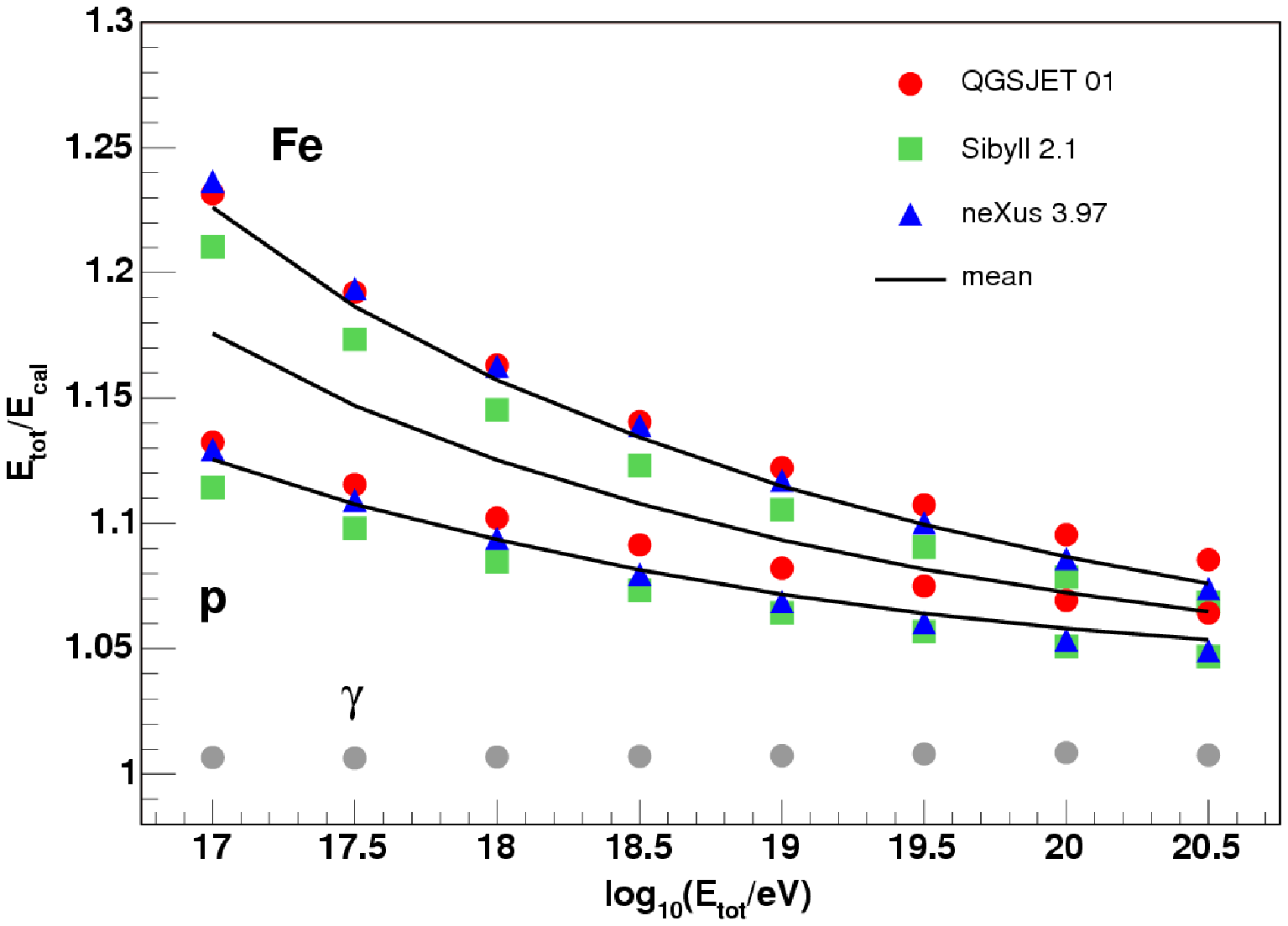}
\caption{\label{figMissE} Correction factor for the missing energy calculated by several models (dots) with fitted curves \cite{missingEnergy} for proton and iron primaries.}
\end{minipage}
\end{figure}

The two measurement techniques in use at the Pierre Auger Observatory -- surface detection and air fluorescence detection -- are quite complementary. (Fig. \ref{figHybridDetection}). Whereas SD has $100\%$ duty cycle time and acceptance can be calculated and it is model independent, FD has duty cycle of $\approx 13\%$ \cite{fdPaper} and acceptance depends on the atmospheric conditions (model dependent). SD measurement corresponds to one slice of a calorimeter at the last stage of shower development. To get the energy only from SD, Monte Carlo interaction models are needed in experiments without fluorescence detectors. On the other side FD collects photons emitted during development of the shower. It enables to sample a shower along its axis. Measured number of photons in different atmospheric slant depths can be converted to deposited energy to get shower longitudinal profile (Fig. \ref{figLogProf}). Then, the energy is measured directly with the atmosphere acting as a calorimeter for CR study. Events that are measured simultaneously by FD and SD are called hybrid.

%Apart from the energy another very important characteristic of the shower development is the position of shower maximum as the atmospheric depth where shower reaches the maximum number of secondary particles are (maximum in shower longitudinal profile).

Apart from the energy another very important characteristic of the shower development is the position of shower maximum. Shower maximum is the atmospheric depth where the shower longitudinal development reaches the maximum number of secondary particles.

%poznamka o tom, ze se zlepsi rekonstrukce

%todo:
%12 procent events with hybrid reconstruction

\subsection{FD Energy Measurement and Shower Reconstruction}
Electromagnetic energy losses of charged particles are responsible for emitted fluorescence light. Therefore FD measures longitudinal shower profile. About $20$ photons are emitted between 300 and 400 nm per 1 MeV loss. The total number of photons is proportional to the deposited energy of the air shower. Thus, the profile integral gives almost the whole shower energy ($\sim 90\%$ of the total energy at $10^{19}$ eV):
\begin{equation}
\label{eqEcal}
E_{cal} = \int dX {dE\over dX}.
\end{equation}
The remaining part ($\sim 10\%$ of the total energy at $10^{19}$ eV) of the shower energy cannot be directly measured by collecting emitted light. It concerns energy of neutrinos, muons and other particles which do not excite nitrogen molecules. This type of energy is called 'missing' and it is not included in the integral of the energy deposit derived from FD data. Nevertheless, one can calculate the correction factor that depends on the primary particle type and energy (Fig~\ref{figMissE}). 

\subsection{SD Energy Calibration}
\begin{figure}[htb]
\begin{minipage}{17.5pc}
\includegraphics[width=17.5pc]{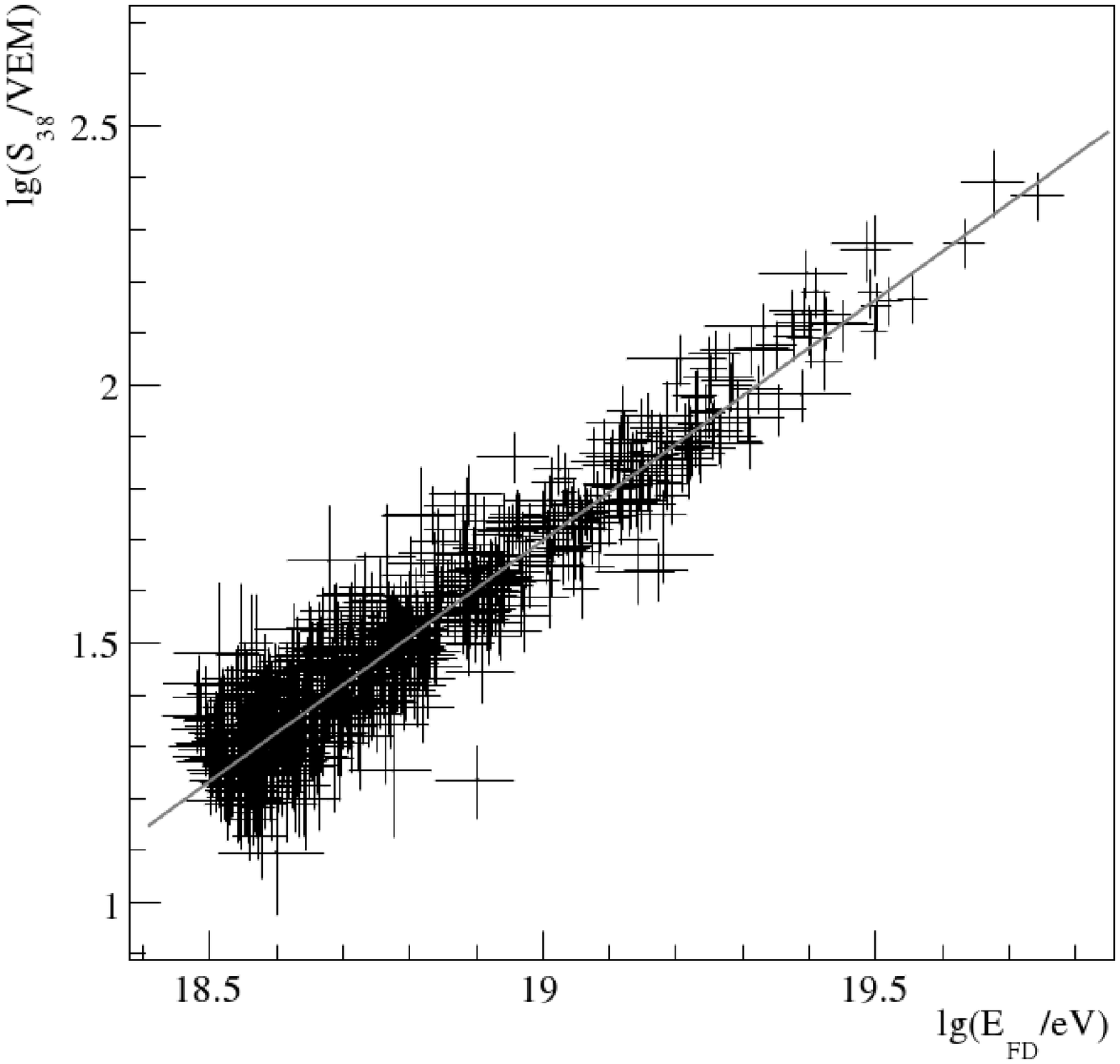}
\caption{\label{figCalib} Ground parameter $S_{38}$ as a function of the energy as measured with the FD for the 795 hybrid events \cite{calib}.}
\end{minipage}\hspace{1pc}%
\begin{minipage}{19.5pc}
\includegraphics[width=19.5pc]{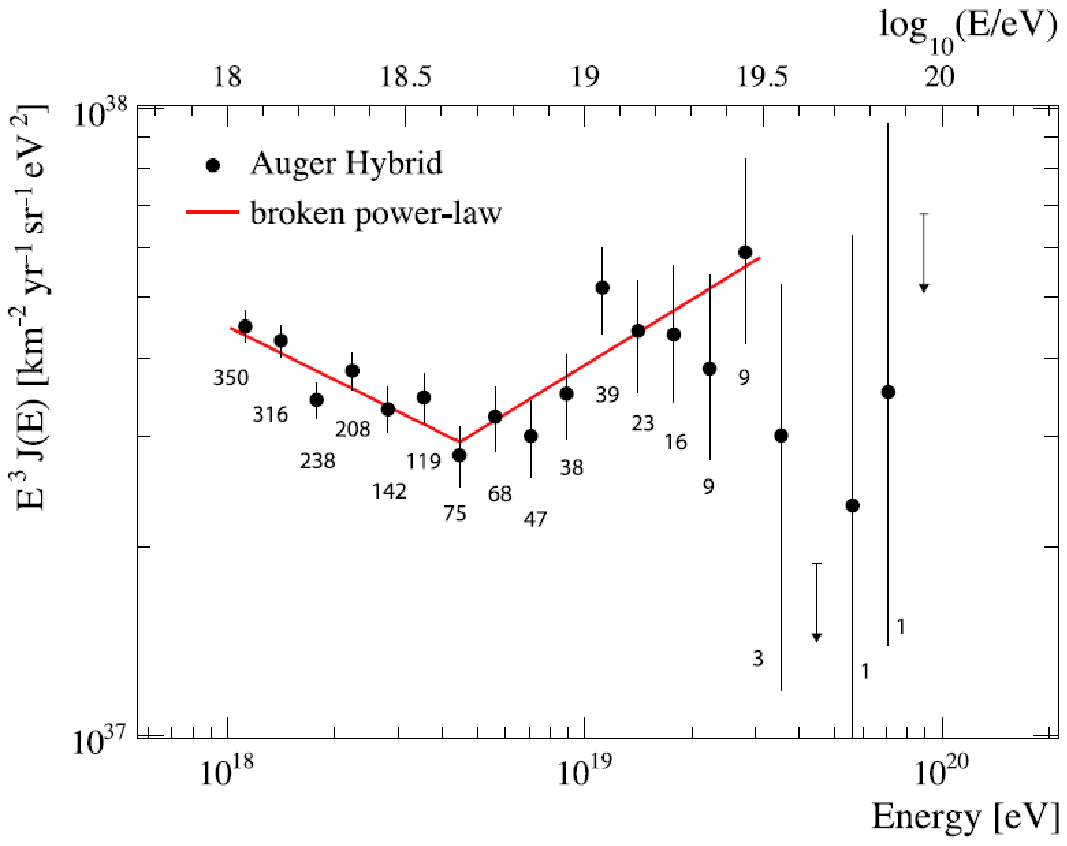}
\caption{\label{figSpectrum} The energy spectrum of UHECR determined from hybrid measurements between November 2005 and May 2008 \cite{spectrum}. Statistical uncertainties are shown with a broken power law used to determine the position of the ankle \cite{spectrum}, which is clearly visible.}
\end{minipage}
\end{figure}

The biggest advantage of hybrid detection is the model independent energy calibration of the SD. The energies of hybrid events are determined from FD measurement, and then the signal from the SD can be calibrated. The SD records signals from electromagnetic and muonic shower component. One shower is typically large enough to hit many SD stations. Therefore one can derive (fit) the lateral distribution function of the signal at ground. The aim of such a procedure is to find the particle signal at a distance of 1000 m from the shower core ($S_{1000}$). The ground parameter $S_{1000}$ depends on the zenith angle of the primary particle. The variable corrected for the zenith angle dependency is called $S_{38}$ and corresponds to signal $S_{1000}$ at $38\degree$ zenith angle. The calibration is based on finding the dependency of $S_{38}$ on the energy measured by FD (Fig.~\ref{figCalib}).

\subsection{Uncertainties on the Reconstructed Energy from FD}
The use of the atmosphere as a calorimeter also determines the main uncertainties of the measurement. 
The total systematic uncertainty in the FD energy scale is $22\%$, with the major contributions due to uncertainties in the fluorescence yield ($14\%$) and atmospheric transmission ($7\%$ for aerosol scattering, $1\%$ for Rayleigh scattering). There are also $9\%$ uncertainties in the telescope absolute calibration, and the longitudinal shower profile reconstruction contributes $10\%$ to the total energy uncertainty. Missing energy gives $4\%$ uncertainty. 

Several devices have been installed to monitor atmospheric conditions, especially those for optical transmission determination. For example, the Central Laser Facility~\cite{clf}, Lidar stations~\cite{lidars}, and the FRAM telescope~\cite{fram} are used for atmospheric monitoring program. 

The measurement of the shower maximum requires a conversion of geometrical altitude as observed by FD to atmospheric depth. The conversion depends on temperature, pressure and humidity, which vary with time and location. There are meteorological stations and a balloon launch facility to measure these thermodynamical parameters. Use of measured conditions at the site considerably improves the accuracy of the air shower reconstruction \cite{atmos}.
%and their knowledge considerably improves EAS shower reconstruction.

The fluorescence yield is the number of photons emitted in a given band per unit of energy loss by charged particles. It is an important quantity for the energy reconstruction. The Pierre Auger Observatory uses the absolute yield from Ref. \cite{nagano}.

%The absolute FY is measured in several experiments, Pierre Auger Observatory uses yield $5.05\pm 0.71$ photons per MeV of energy deposited in air at 293 K and 1013 hPa from 337 nm band \cite{nagano}. FY is also affected by atmospheric parameters (pressure, humidity, temperature). The wavelength dependence (relative ratio between yields at different bands) is taken from \cite{mAve} (Fig. \ref{figFluorYield}). 

\section{Conclusions}
The fluorescence detector of the Pierre Auger Observatory is an atmospheric calorimeter used for the hybrid detection of air showers. It enables a model-independent energy measurement and calibration of surface detector array. The first data taking started in late 2003. The Observatory has operated steadily at its full design size since July 2008. The fluorescence detector data have enabled for example precise measurements of the UHECR energy spectrum (Fig. \ref{figSpectrum}), elongation rate \cite{rate} and CR composition.

\section{Acknowledgements}
The contribution is prepared with the support of Ministry of Education, Youth and Sports of the Czech Republic within the project LA08016 and with the support of the Charles University in Prague within the project 119810. Special thanks to my colleages from the Pierre Auger Collaboration for fruitful cooperation.

\end{document}